# Implicit Neural Representations Framework for One-Dimensional Magnetotelluric Inversion

**Fareeda Begum Shaik[1], Roshan K. Singh [1 *], Pankaj K Mishra [2]**

[1]Indian Institute of Petroleum and Energy, Visakhapatnam, India

[2]Geological Survey of Finland, Vuorimiehentie 5, Espoo, Finland

*Corresponding author: Roshan K Singh IIPE

---

**Abstract:**

Magnetotelluric (MT) inversion is a very useful technique to image the subsurface electrical resistivity structures. It is used for mineral exploration, geothermal studies, groundwater assessment, and lithospheric investigations. In this work, we proposed a physics-informed machine learning framework for 1D MT inversion based on implicit neural representations (INR). Our approach models the subsurface resistivity as a continuous function of depth using a coordinate-based neural network. This method does not require fixed discretization or layered models. The neural network is trained directly on a differentiable MT forward-model loss based on Wait's recursive impedance formulation. This setup allows inversion to occur in a physics-consistent optimization framework. The implicit regularization avoids the need for manual tuning of external regularization. We have tested this method on synthetic conductor models and real MT data. The results showed its ability to recover geologically relevant resistivity structures over various depths and thicknesses. Through different initializations, we can compute an ensemble of plausible models to estimate model uncertainty. These results suggest that implicit neural representations provide a flexible framework for geophysical inversion, with even greater potential in higher-dimensional MT problems and joint inversion applications.

---

## 1. Introduction:

Magnetotelluric (MT) methods are generally used to map subsurface electrical resistivity structures. They can be applied to a range of spatial scales, from near-surface studies to deep crust and mantle studies. The method exploits naturally occurring electromagnetic fields generated by ionospheric and magnetospheric current systems and provides information about subsurface conductivity through the frequency-dependent impedance relationship between electric and magnetic fields. Estimating the resistivity structure from MT data requires solving a nonlinear inverse problem, which is inherently ill-posed and non-unique (Kang et al., 2017). The sensitivity of MT responses generally decreases with depth due to the diffusive nature of electromagnetic fields and the inherent physical laws governing the MT method. Furthermore, MT inversion is an ill-posed and non-unique problem, meaning that the number of knowns is less than the number of unknowns and different resistivity models can produce similar apparent resistivity and phase responses (Simpson and Bahr, 2005). Therefore, classical inversion approaches employ stabilizing constraints or regularization techniques to obtain geologically plausible and stable solutions. However, inversion methods differ in their ease of implementation and in the choices of model parameterization, regularization strategy, initial model selection, and optimization scheme used during the inversion process. In addition, the type and amount of information recovered from the inversion may also vary, and most 1-D inversion methods utilize one of the apparent resistivity modes, with or without the corresponding phase information (Nabighian, 1991).

One of the most widely used deterministic inversion approaches is Occam inversion, introduced by (Constable et al., 1987). The method seeks the smoothest resistivity model that adequately fits the observed data and has been extensively applied in MT studies due to its stability and robustness (De Groot-Hedlin and Constable, 1990; Key, 2009). However, Occam inversion may suppress sharp resistivity contrasts associated with geological structures such as faults, intrusive bodies, and fluid pathways because of the imposed smoothness constraints (Vargas Huitzil et al., 2025). Besides deterministic approaches, several stochastic and metaheuristic inversion techniques have also been developed for MT applications. For example, simulated annealing (Shaw and Srivastava, 2007), particle swarm optimization, and genetic algorithm-based inversion methods (Chandra et al., 2017) have been applied to solve complex and multi-modal solution spaces more effectively. Bayesian (Guo et al., 2011) and Markov Chain Monte Carlo (MCMC)-based inversion approaches (Mosegaard and Tarantola, 1995) are also widely used to quantify model uncertainty and non-uniqueness. Furthermore, regularized nonlinear conjugate gradient (Newman and Alumbaugh, 2000) and data-space inversion methods (Siripunvaraporn and Egbert, 2000) have gained significant attention in large-scale geophysical inversion problems due to their computational efficiency and improved model recovery.

Deep learning (DL) is a subfield of machine learning (ML) based on multi-layer neural networks and has demonstrated remarkable ability in modeling complex nonlinear relations in large geophysical data sets. The recent advances in ML and scientific computing have triggered the application of neural networks to the geophysical inversion problems. Some deep learning methods, such as two-dimensional deep learning inversion, physics-driven deep neural network (PhyDNN), convolutional neural network (CNN)-based MT inversion (Liu et al., 2021), and transformer-based MT forward modeling and inversion (Deng et al., 2023), have been developed to learn the mapping from geophysical observations to subsurface resistivity structures. DL-based inversion methods do not require the recomputation of the sensitivity matrix, which describes the relationship between the perturbations of model parameters and the corresponding changes of the observed data, in contrast to the conventional inversion approaches. Also, these methods are often less sensitive to the initial model choice (Xu et al., 2024). However, data-driven approaches require large synthetic training datasets and often do not generalize well to complex real-world geological environments. Therefore, recent studies have been dedicated to physics-informed or physics-driven learning approaches where the governing physical laws, such as Maxwell's equations, and boundary conditions are embedded in the training of neural networks to combine the domain knowledge and machine learning to solve geophysical inverse problems (Goyes-Peñafiel et al., 2025).

An implicit neural representation (INR) (also called a coordinate-based network or neural field), which encodes a continuous mapping from spatial coordinates to property (resistivity) values (Sitzmann et al., 2020) is an alternative paradigm. INRs have demonstrated remarkable success in computer graphics for representing images and three-dimensional scenes (Essakine et al., 2024) and have recently been used in earth science applications such as seismic inversion (Romero et al., 2025; Sun et al., 2023), potential-field data processing (Smith et al., 2025), geological modeling (Hillier et al., 2023), gravity-field modeling (Izzo and Gómez, 2022; Schuhmacher et al., 2023), synthetic 2D inversions of DC and seismic data (Xu and Heagy, 2025), and a similar kind of approach on 3D gravity inversion (Li et al., 2026). More recently, INRs have been explored in geophysics for seismic inversion, potential-field inversion, and electrical resistivity problems. For example, a recent study introduced a neural-field representation for three-dimensional gravity inversion, demonstrating that subsurface density structures can be recovered by training a neural network directly against the gravity forward model (Mishra et al., 2025). In this study, we extend the INR framework to one-dimensional MT inversion, where the subsurface resistivity distribution is represented as a continuous neural function of depth. The neural network is trained directly using a differentiable MT forward modeling operator implemented within an automatic differentiation framework. The physically guided deep unsupervised inversion (Goyes-Peñafiel et al., 2025) combines a neural inverse operator with a differentiable MT forward model, which removes the need for labeled datasets but still requires a predefined layered parameterization and explicit Tikhonov regularization. In contrast, the INR method assumes that the subsurface is a continuous function of resistivity with respect to the coordinate. This method eliminates the need for a priori defined layer interfaces and allows the estimation of resistivity at arbitrary depths. Moreover, the model discretization is decoupled from the number of trainable parameters by INR, allowing for higher-resolution representations without an accompanying increase in parameters. In addition, the neural parameterization provides an implicit regularization via architectural bias, which reduces the need for explicitly designed regularization constraints. The proposed approach also preserves fine-scale conductivity variations and minimizes staircase artifacts characteristic of discrete layered models (Mishra et al., 2025). Thus, the use of INR gives a more flexible and geologically realistic MT inversion framework.

The primary objectives of this work are to explore the capabilities of the implicit neural representation (INR) framework for the retrieval of subsurface resistivity structures from MT data, to analyze the effect of network architecture and depth parameterization on the performance of the inversion, and to compare the performance of INR-based inversion with the conventional Occam inversion approach using synthetic and real MT datasets. In addition, the proposed method is tested by synthetic conductor tests and various performance metrics to prove its reliability and robustness.

## 2. Methodology:

### 2.1 Forward modeling of Magnetotelluric Responses

#### 2.1.1. 1D Magnetotelluric Forward Modeling Using Wait's Recursive Formulation

The magnetotelluric (MT) responses were calculated using the classical recursive impedance formulation for a 1-D layered Earth (Wait, 1982). The subsurface is modeled as a stack of horizontal layers with specified resistivity and thickness over a homogeneous half-space representing the deep earth (Fig. 1). The frequency domain impedance is calculated recursively, propagating the impedance from the bottom layer to the surface, by use of Wait's recursive relation, which is derived from Maxwell's equations. This method is an efficient and stable solution for one-dimensional layered Earth models. The surface impedance so obtained is then used to calculate the apparent resistivity and phase responses, which describe the amplitude and phase relationship between the surface electric and magnetic fields. This methodology is widely used in MT forward modeling and inversion (e.g., Grandis, 1999; Rokityansky, 1982; Pek and Santos, 2006).

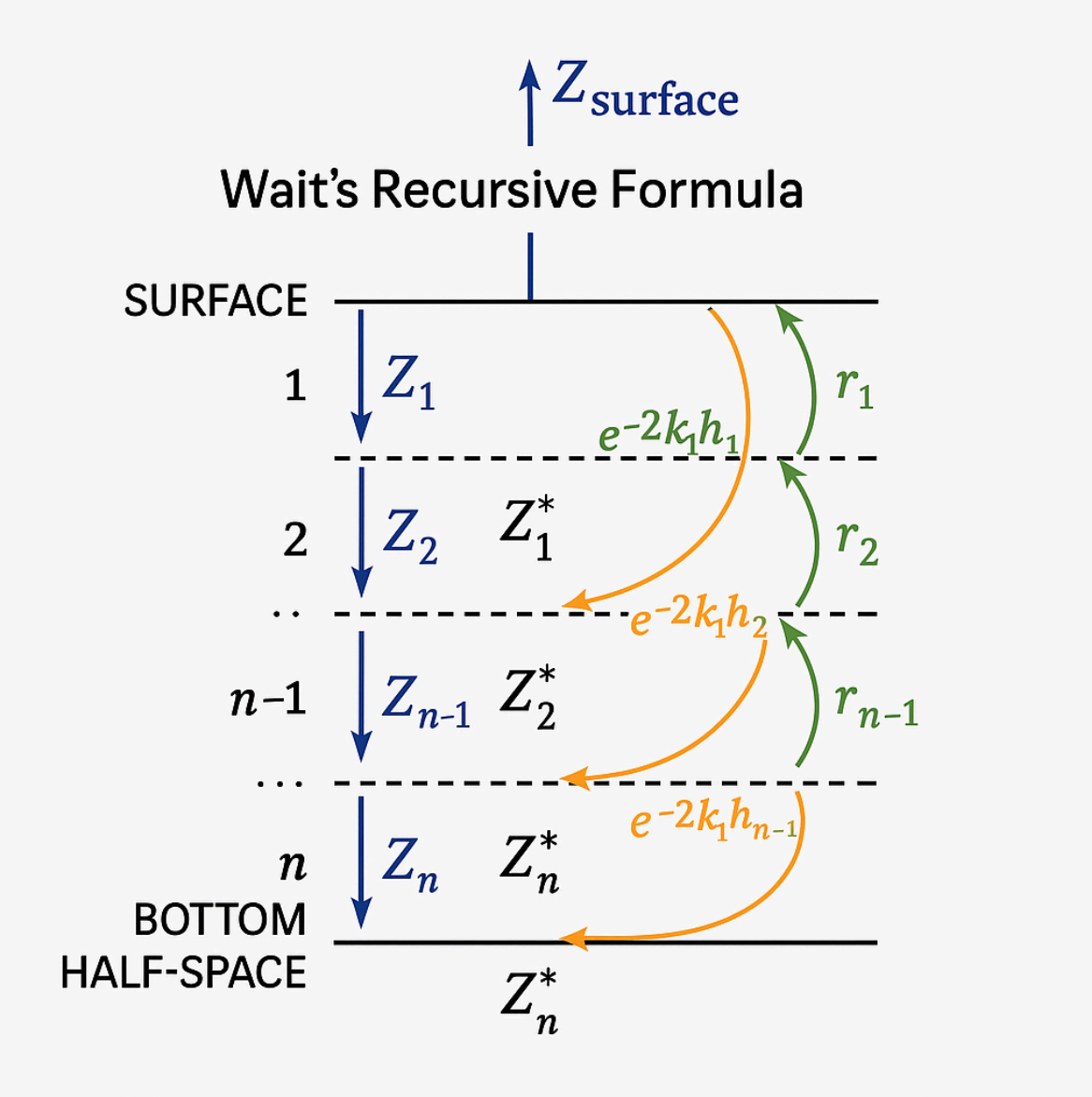

*Fig. 1: Wait's recursive impedance formulation for one-dimensional MT forward modeling in layered media.*

**2.1.2. Differentiable Implementation for Neural Inversion:** In this study, the recursive impedance equations are applied using the PyTorch automatic differentiation framework (Kordowich and Jaeger, 2025). It enables the forward modeling procedure to be fully differentiable. The impedance recursion and the calculation of apparent resistivity and phase are indicated using complex-valued tensor operations.

Embedding the forward solver within a differentiable computational graph allows gradients of the predicted MT responses with respect to neural network parameters to be obtained automatically via backpropagation (Waldo, 2022). This enables the implicit neural representation model to be trained directly using the physics-based forward operator without explicit derivative calculations. The computational efficiency and differentiability of Wait's recursion make it particularly suitable for integration with physics-based neural inversion frameworks.

## 2.2. Implicit Neural Representation of the Resistivity Model

In this study, the subsurface resistivity distribution is represented using an implicit neural representation (INR), also known as a neural field or coordinate-based neural network. Instead of estimating resistivity values for discrete layers directly, the subsurface resistivity is modeled as a continuous function of depth $\rho(z) = f_\Theta(z)$, where $f_\Theta$ is a neural network parameterized (Xu and Heagy, 2025) by weights θ. To improve numerical stability and ensure positivity of resistivity, the neural network predicts log-resistivity $\log \rho(z) = f_\Theta(z)$ (Fig. 2). The resistivity model is therefore obtained as $\rho(z) = \exp(f_\Theta(z))$

The neural network used in this study is a multilayer perceptron (MLP) consisting of several fully connected layers with nonlinear activation functions. Each hidden layer uses the LeakyReLU activation function, while the output layer uses sigmoid or Tanh activation functions to constrain the predicted log resistivity (Fig. 2) within predefined bounds. The output of the neural network is mapped to the physical resistivity range $\log \rho_{min} \leq \log \rho_z \leq \log \rho_{max}$ ensuring that predicted resistivity values remain geophysically realistic.

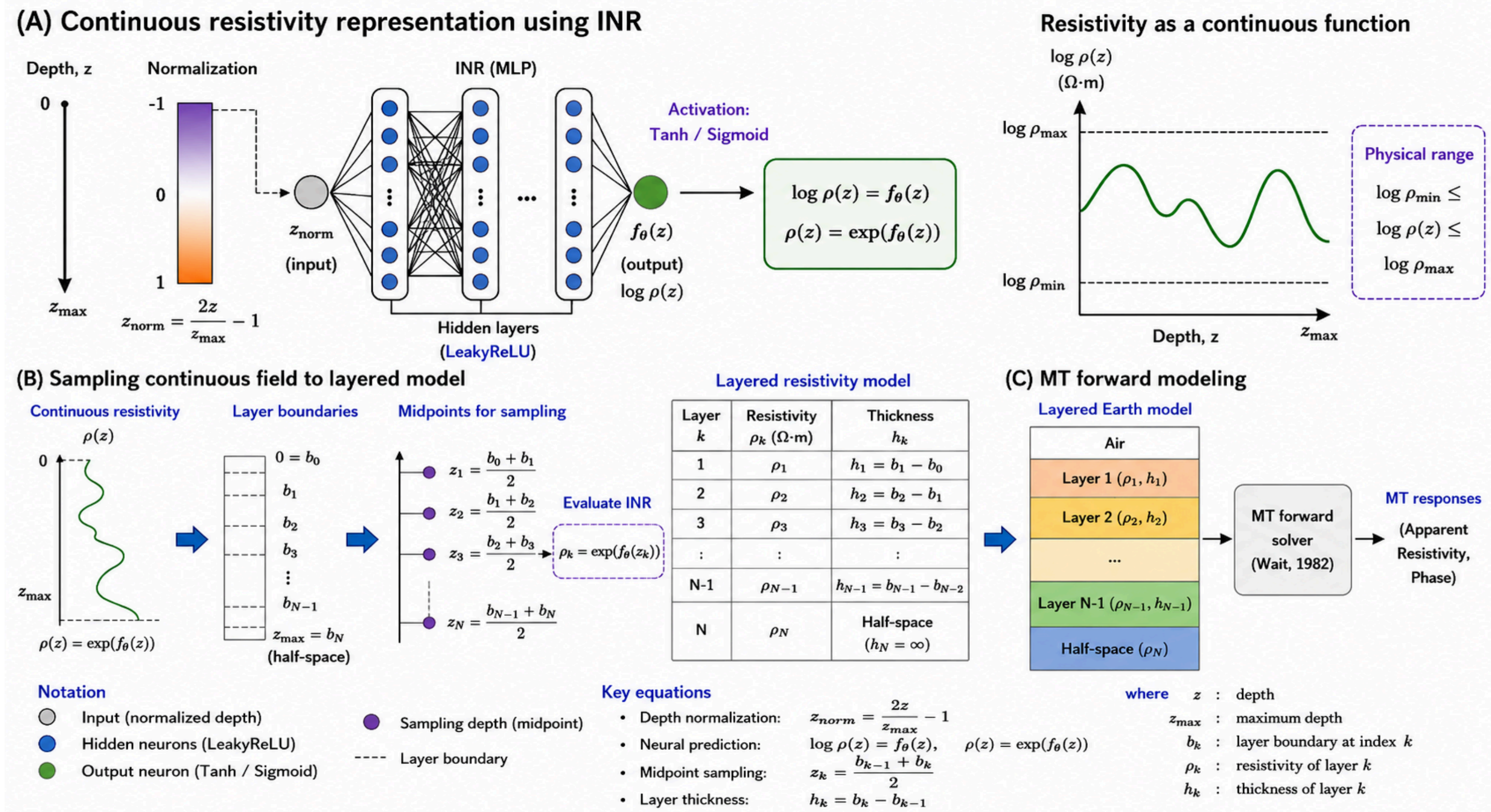

*Fig. 2: Schematic overview of the INR-based resistivity parameterization and its integration with layered forward modeling*

The depth coordinates are normalized by mapping them to the range [−1, 1] using $z_{norm} = (2z/z_{max}) - 1$ to stabilize neural network training. Although the INR represents resistivity as a continuous function of depth, the magnetotelluric forward solver requires a layered Earth parameterization. To bridge this gap, the continuous neural resistivity field is sampled onto a discretized layered structure before computing MT responses. The model domain is divided into $N$ layers defined by depth boundaries, $b_1$, $b_2$, $b_3$,....$b_{N-1}$. The resistivity of each layer is obtained by evaluating the neural field at the midpoint of each layer, $z_k = (b_{k-1} + b_k)/2$. The neural network is evaluated at these depths to obtain resistivity of each layer, $\rho_k = \exp(f_\Theta(z_k))$. Layer thicknesses are computed as $h_k = b_k - b_{k-1}$ with the deepest layer treated as a half-space extending to infinite depth. This sampling procedure enables the continuous neural representation to be interfaced with the classical layered MT forward modeling algorithm while preserving the flexibility of the neural field.

## 3. Results and Discussion

### 3.1. Synthetic Case

A synthetic five-layer resistivity model (Fig. 3) from a benchmark nonlinear MT inversion study (Kang et al., 2017). To mimic realistic field observations, we added stochastic noise to the synthetic MT responses. For apparent resistivity, a log-normal distribution, and for phase, a Gaussian distribution. The inversion included measurement uncertainties by using a weighted least squares formulation, with the weighting factors being the standard deviations of the apparent resistivity and phase. This formulation can cope with data errors in a more realistic way than traditional unweighted mean square error (MSE) approaches.

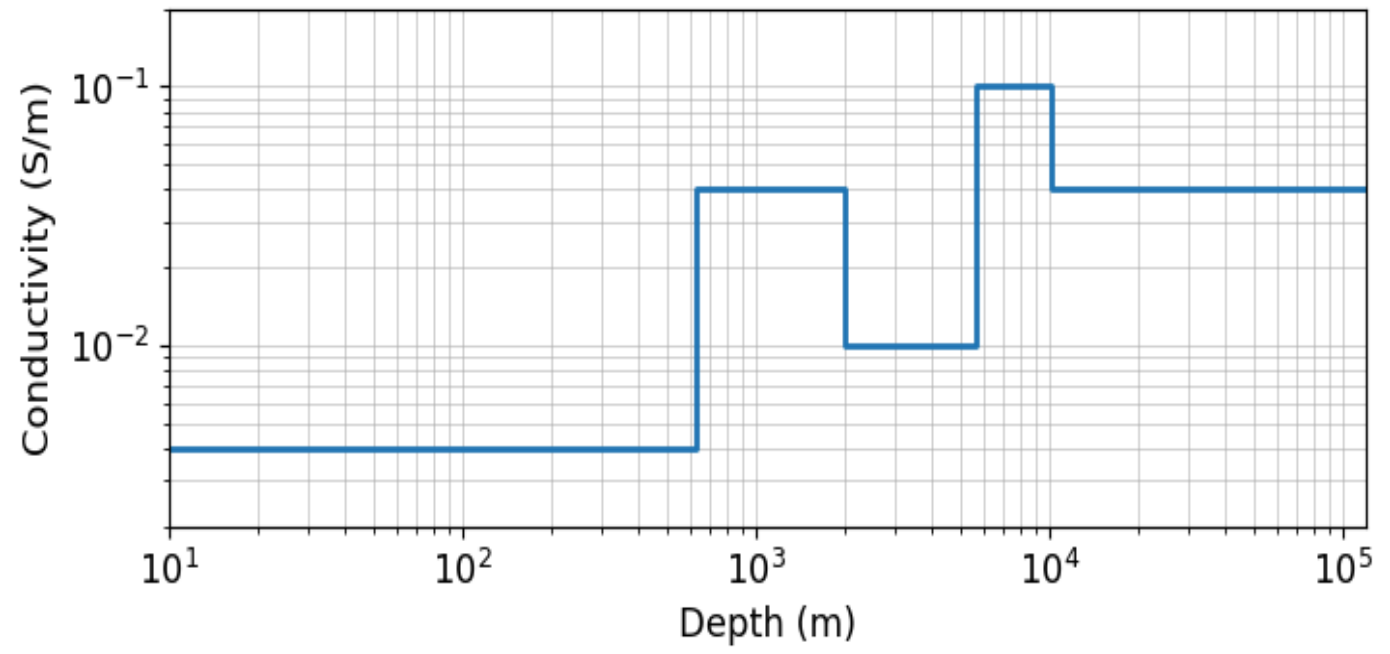

*Fig. 3. Synthetic true model (Kang et al., 2017).*

| Resistivity (Ω·m) | Thickness (m) | Depth (m) |
|---|---|---|
| 250 | 600 | 600 |
| 25 | 1391 | 1991 |
| 100 | 3795 | 5786 |
| 10 | 4000 | 9736 |
| 25 | Half-space | infinite |

*Table 1. Synthetic true model values.*

#### 3.1.1. Network Architecture Sensitivity Analysis

Unlike conventional inversion algorithms, the INR framework does not require an explicit mesh-based model parameterization (Mishra et al., 2025). The influence of neural network architecture on reconstructed resistivity models and respective MT responses is shown in Figure 4 by varying the number of hidden layers and neurons. Our results indicate that smaller architectures such as shallow networks with fewer neurons (Fig. 4a-b) are not sufficient to capture the variations of the resistivity in the subsurface. These networks show underfitting behavior and poor recovery of the conductive layer geometry with large deviations of the apparent resistivity and phase fits. Their limited representational capacity does not allow them to capture complex variations of the resistivity field. Both the recovered model and MT responses are significantly improved by increasing the network size (Fig. 4c–g). Architectures of moderate size are more flexible in representing the continuous resistivity function, which results in a better recovery of the conductive structure and a better fit to the observed apparent resistivity and phase curves. The recovered models are smoother and preserve the main resistivity contrasts. Inversion results for larger architectures with more hidden layers and neurons (Fig. 4h–j) show only marginal improvement over the moderate-sized networks. The fit of the data remains good, but the resistivity models become increasingly smooth, and the computational complexity increases due to the larger number of trainable parameters. The network with 4 hidden layers and starts with 128 neurons in the first hidden layer (Fig. 4d) provides the best trade-off between model recovery, quality of data fit, and computational efficiency among the tested architectures. This architecture accurately reproduces both apparent resistivity and phase responses while recovering the main resistivity structure without unnecessary model complexity.

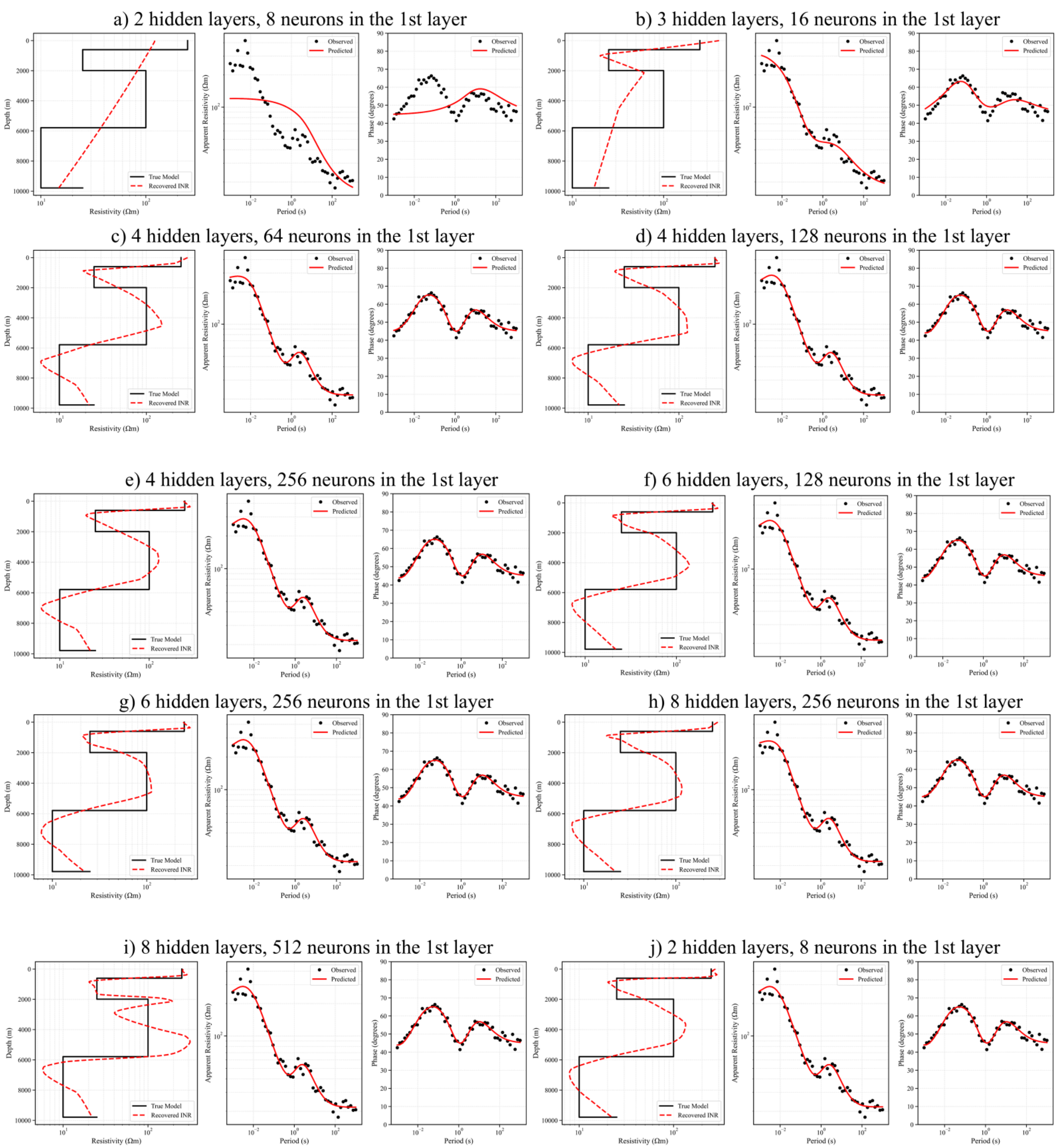

*Fig. 4: Examination of the effect of neural network size and depth on MT inversion responses using different INR architectures. a)-b) a small-sized neural network, b)-g) a moderate-sized neural network, and h)-j) a large-sized neural network*

### 3.1.2. Principle of Equivalence Test

The principle of equivalence and the inherent non-uniqueness of MT inversion were tested by using two synthetic models (Fig. 5, left panel). The two different layered resistivity models (Model A and Model B) shown in (Fig. 5, center panel) (data space) generate nearly similar apparent resistivity responses despite their different subsurface structures. This ambiguity is due to the diffusive nature of the electromagnetic fields in MT, which are less sensitive at depth and are determined by the integrated conductivity distribution rather than sharp resistivity boundaries. Thus, different subsurface models with similar conductivity-thickness products (conductance) can give identical MT responses. In this case, Model A is made of a thinner, more conductive layer, and Model B is made of a thicker, less conductive layer. Although resistivity and thickness values differ, their effective conductance is comparable, leading to similar MT responses. Fig. 5, right panel (model space), compares the two equivalent true models with the INR-recovered resistivity profile. The recovered INR model does not exactly reproduce either Model A or Model B, but the inversion converges toward a smooth and continuous resistivity distribution representing the effective conductivity structure. The conductive anomaly near ~2000–3000 m depth is identified, although it appears broader and smoother than the true layered models. This result indicates that the INR framework provides a data-consistent solution within the equivalence class rather than forcing the inversion toward a unique layered model.

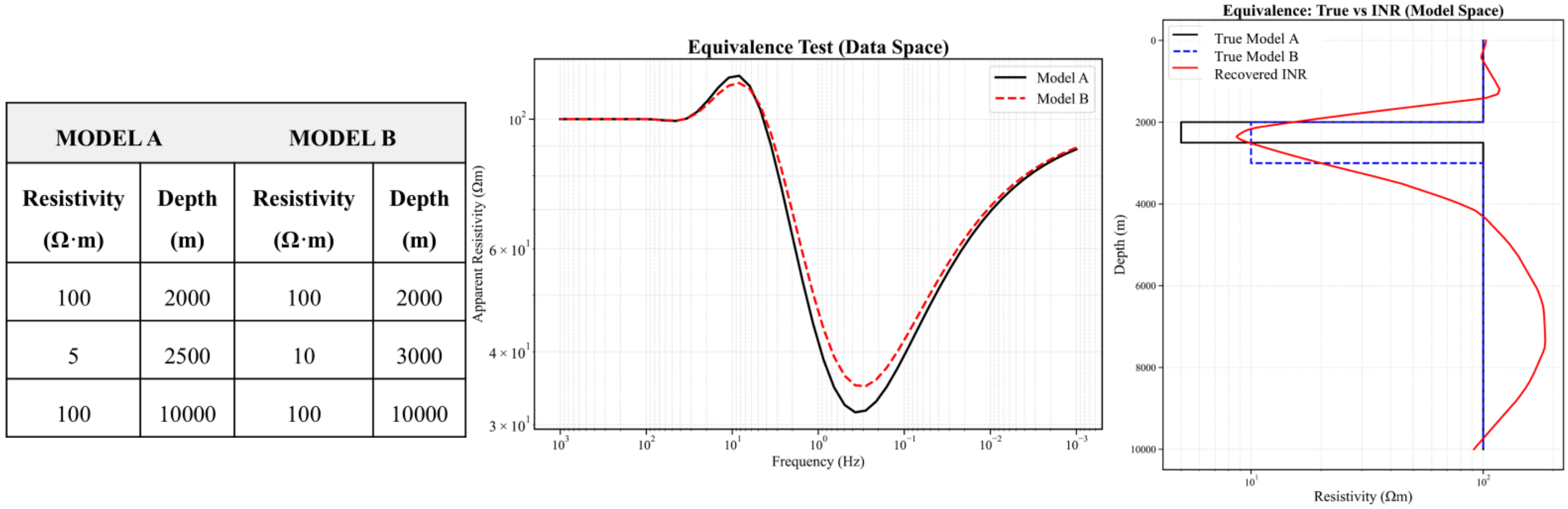

| MODEL A | | MODEL B | |
|---|---|---|---|
| Resistivity (Ω·m) | Depth (m) | Resistivity (Ω·m) | Depth (m) |
| 100 | 2000 | 100 | 2000 |
| 5 | 2500 | 10 | 3000 |
| 100 | 10000 | 100 | 10000 |

Equivalence: True vs INR (Model Space)

*Fig. 5. An equivalence principle test was performed to assess the performance of INR in addressing non-uniqueness. The left panel shows the table of the apparent resistivity and depth values of Model A and Model B; the center panel shows nearly similar MT apparent resistivity responses for two different layered resistivity models, while the right panel compares the equivalent true models with the INR-recovered continuous resistivity model.*

### 3.1.3. Single Conductor Recovery Test

To evaluate the robustness of the INR framework using a series of single-conductor tests at varying depths (D) and thicknesses (T). Figure 6 shows the recovery of the conductive bodies with different burial depths (D=100-8000 m) and thicknesses (T=50-1000 m) with the INR framework. The results show that the quality of the recovery is dependent on burial depth and the conductor thickness. For shallow conductors (100–500 m depth), the conductive anomaly is recoverable for all thicknesses tested. Nevertheless, thin conductors (50 m) are recovered as broadened anomalies while thick conductors (200-1000 m) show better geometric agreement with the true model. At intermediate depths (~2000 m), the recovery starts to deteriorate for thinner conductors. The recovered models are smoother, and the boundaries of the conductor are not as well defined, although the location of the anomaly can still be found. In contrast, conductors of thickness 500-1000 m are reasonably resolved. For deeper conductors (5000-8000 m) the limitations of the method are more pronounced. Thin conductors (50-200 m) are only poorly recovered and are seen as diffuse conductivity perturbations rather than sharply delineated bodies. The recovered models for thicker conductors show an increase in smoothing and spreading vertically with depth. This behavior is caused by the decreased sensitivity of the MT responses to deep targets and the decreasing resolution of the structures with small conductivity-thickness products. In general, the results show that the INR framework can detect conductive structures in a broad range of depths, but its resolving power decreases progressively with depth and for thinner targets. The best recovery is for conductors with larger conductance (higher conductivity-thickness products), while deep, thin conductors remain difficult to resolve.

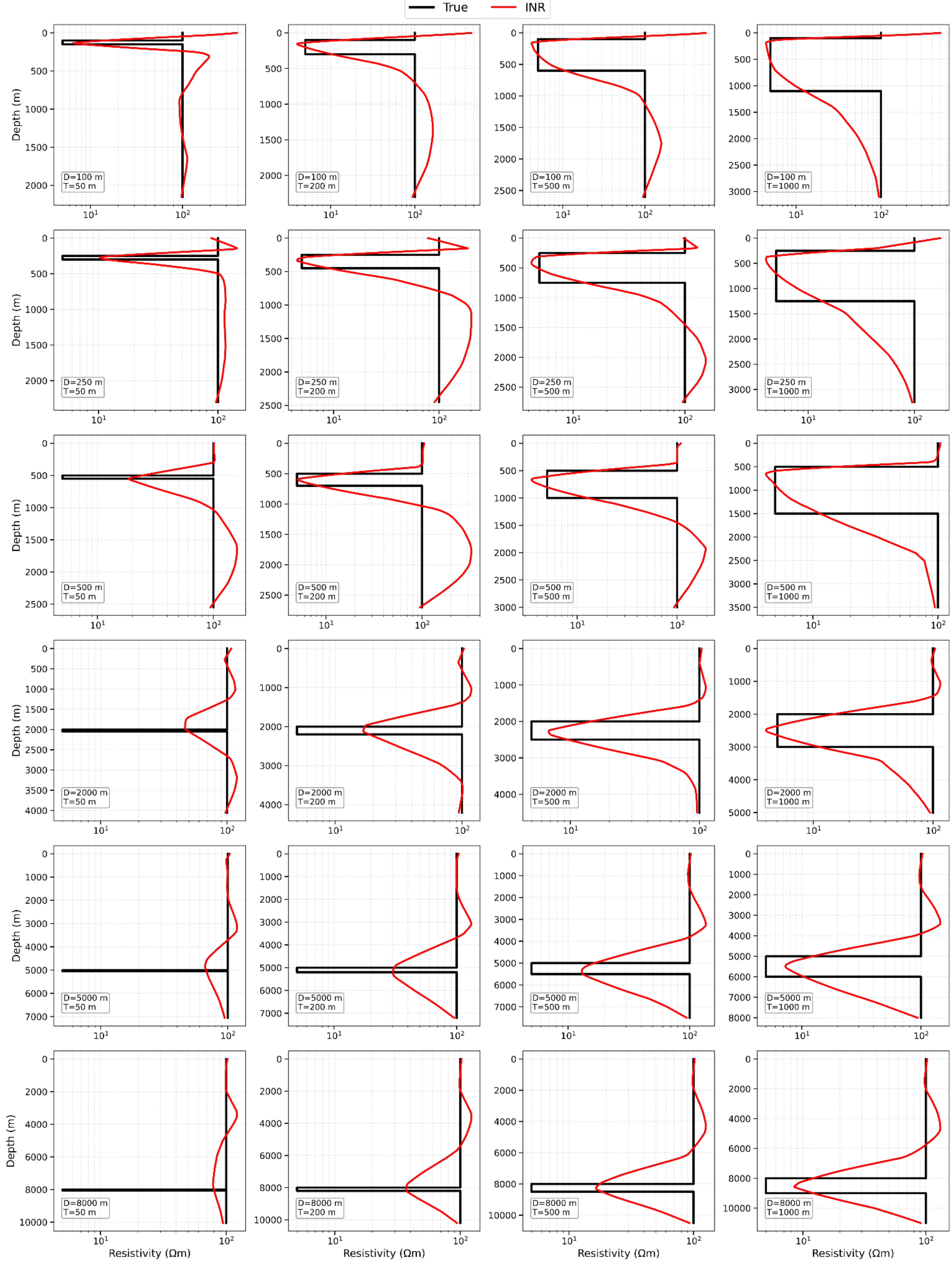

True
INR
Depth (m)
Resistivity (Ωm)
D=100 m T=50 m
D=100 m T=200 m
D=100 m T=500 m
D=100 m T=1000 m
D=250 m T=50 m
D=250 m T=200 m
D=250 m T=500 m
D=250 m T=1000 m
D=500 m T=50 m
D=500 m T=200 m
D=500 m T=500 m
D=500 m T=1000 m
D=2000 m T=50 m
D=2000 m T=200 m
D=2000 m T=500 m
D=2000 m T=1000 m
D=5000 m T=50 m
D=5000 m T=200 m
D=5000 m T=500 m
D=5000 m T=1000 m
D=8000 m T=50 m
D=8000 m T=200 m
D=8000 m T=500 m
D=8000 m T=1000 m

*Fig. 6. Single conductor recovery tests with varying depths and thicknesses. The results show the effectiveness of the INR framework in reconstructing conductive structures and highlight the reduction in resolution.*

### 3.1.4. Uncertainty Quantification and RMS Convergence Behavior

To further examine the stability of the inversion results, an uncertainty quantification (UQ) was studied using an ensemble approach with 40 different random initialization seeds (Fig. 7, right). The ensemble responses generated are close to the true model and show low variability, suggesting that the INR inversion is relatively robust to random initialization of the network. We also examined the convergence behavior of the RMS with an RMS target criterion (Fig. 7, left). The inversion was stopped when the RMS reached a target value of 1.0 to avoid overfitting while achieving an adequate fit to the observed data. The results show that the proposed INR framework can generate reproducible and stable inversion results and is not overly sensitive to the choice of initial network parameters.

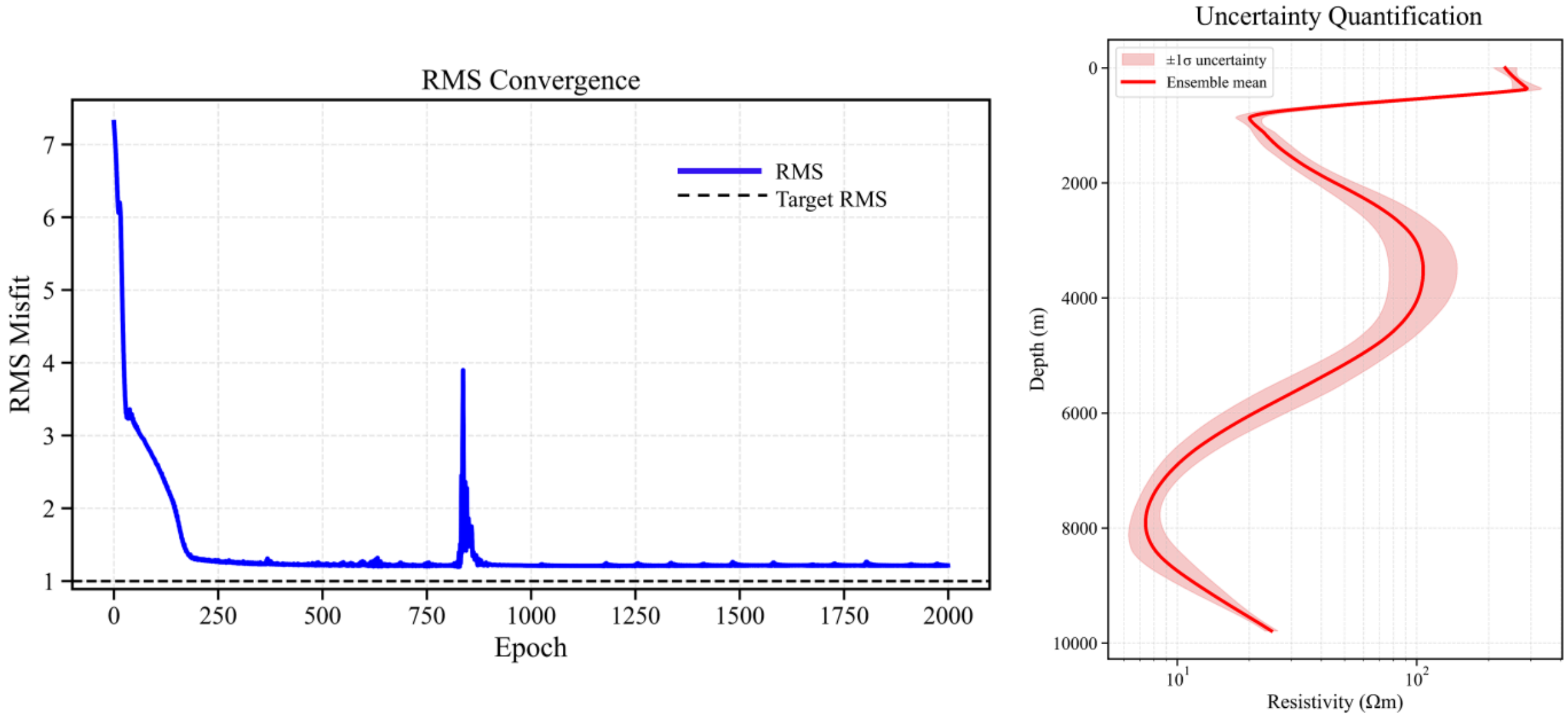

*Fig. 7. Uncertainty quantification and RMS convergence analysis for the INR-based MT inversion framework. The left panel shows the RMS convergence behavior toward the target misfit value during training, while the right panel presents the ensemble inversion results obtained using 40 different random initialization seeds, demonstrating the stability and reproducibility of the proposed approach.*

### 3.1.5. Data fits and Model Comparison between INR and Occam inversion

The data-fitting performance and recovered resistivity structures of the proposed INR framework were compared with the traditional Occam inversion method . The comparison shows that the INR approach gives a better agreement with the observed MT responses than the Occam inversion, especially in the apparent resistivity and phase responses (Fig. 8, left & center). This conclusion is drawn on the basis of smaller deviations between the observed and predicted curves obtained by the INR model over the investigated period range. The Occam inversion stabilizes the solution by explicit smoothness regularization, which usually leads to smooth resistivity distributions but may reduce the sensitivity to local changes in the subsurface model. By contrast, the INR framework considers resistivity as a continuous neural function and optimizes the model with all available MT observations simultaneously. The neural parameterization gives implicit regularization through the architecture of the network so that a stable inversion is possible without explicitly enforcing smoothness constraints. The INR-recovered model (Fig. 8, right) is a better representation of the main conductive and resistive features of the true subsurface than the Occam result, while it is smooth and continuous. The results indicate that the INR framework can better balance data fit and model recovery without explicit regularization parameters.

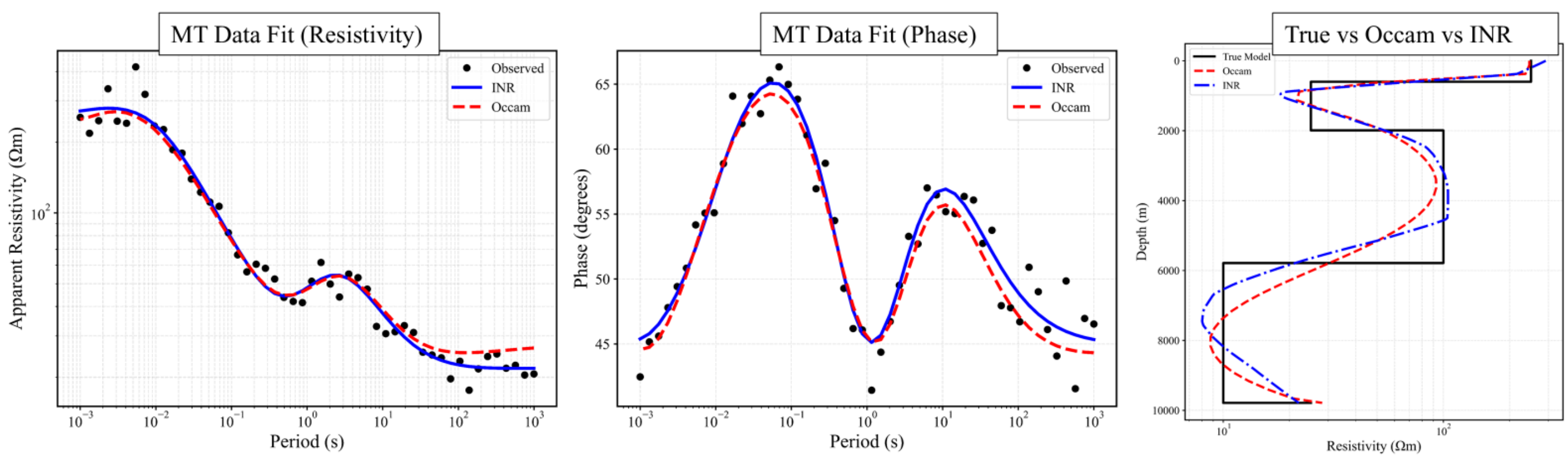

*Fig. 8. Comparison of MT data fitting and recovered resistivity models obtained using the INR and Occam inversion methods. The left and center panels compare the observed MT responses with the responses predicted by the two inversion approaches for apparent resistivity and phase. The right panel compares the corresponding recovered resistivity models with the true subsurface model*

### 3.1.6. Regularization Sensitivity Analysis of Occam and INR

The Occam inversion is very sensitive to the choice of the regularization strength. This dependence is demonstrated in the regularization study shown in Fig. 9, where different regularization weights are tested in the ranges λ = 1–50 and λ = 10–100. Large variations are observed in the resistivity models obtained for different regularization weights in the Occam results (Fig. 9, left & center). The smaller the regularization, the more unstable the model, with more pronounced fluctuations, and the larger the regularization, the smoother the response. This implies that the derived Occam model is sensitive to the choice of smoothing parameter, and the inversion result is strongly dependent on the applied regularization strength. The INR results (Fig. 9, right), on the other hand, show only small differences for different smoothing values. Despite the variations of the regularization parameter, the recovered resistivity profiles are almost identical and preserve the main subsurface structure. We find that the INR framework is less sensitive to explicit smoothing constraints, as the neural representation itself provides an implicit regularizer through architectural bias. This makes the inversion stable, and a constant model response is achieved. Reduced dependence on externally selected regularization parameters enhances the robustness and reliability of the INR inversion framework.

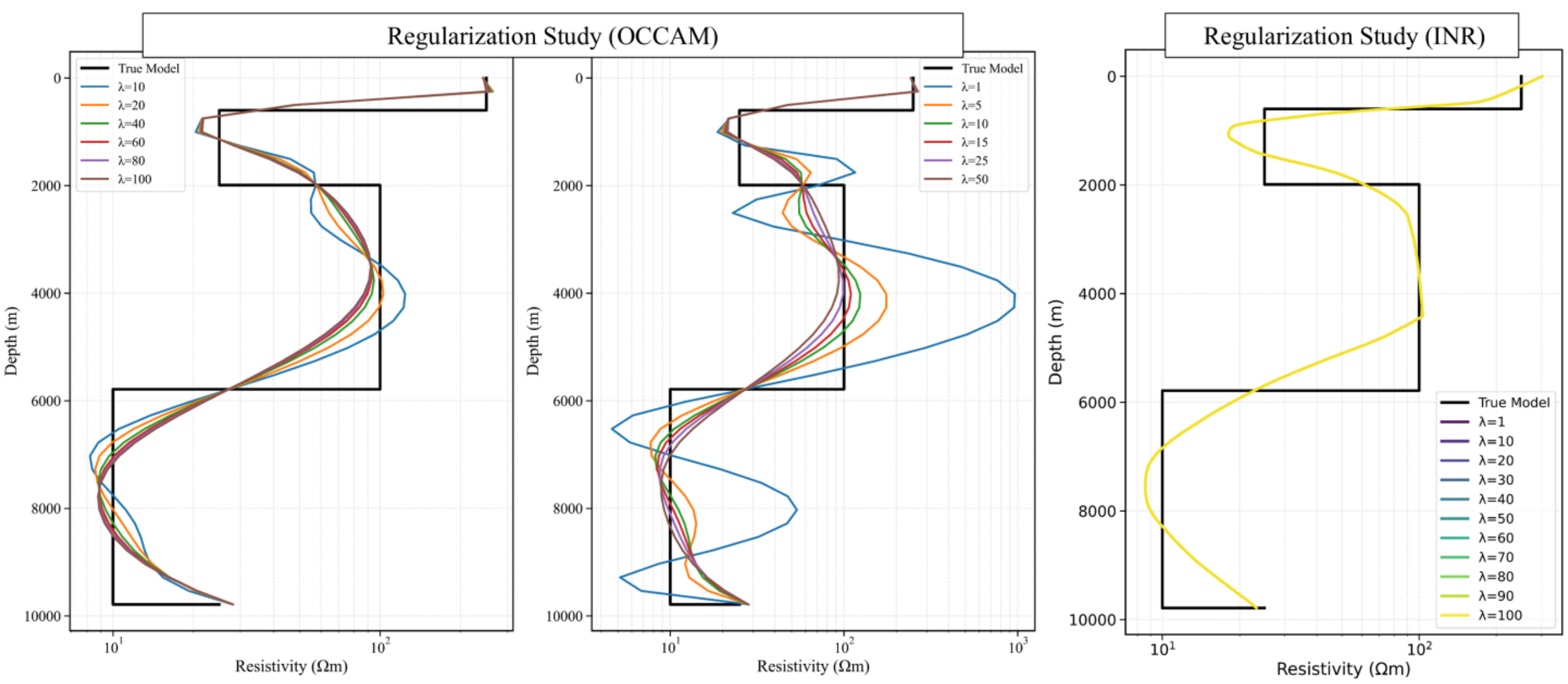

*Fig. 9. Regularization study comparing the sensitivity of Occam and INR inversion approaches. The left and center panels show Occam inversion results obtained using different regularization weights, illustrating the strong dependence of the recovered models on*

*the choice of smoothing parameter. The right panel shows the results of the INR inversion for different smoothness values. The model was relatively stable due to the implicit regularization of the neural network architecture.*

### 3.1.7. Residual Comparison of INR and Occam Inversion

To further assess the inversion quality, normalized residuals for both the Occam and INR methods were computed (Fig. 10). The residual analysis reveals a noticeable peak in the apparent resistivity residuals of the INR inversion at shallow depths. At larger depths, however, the INR residuals tend to be smaller than those of the Occam inversion, indicating that the data are better fitted and the model is more stable in the deeper parts of the subsurface.

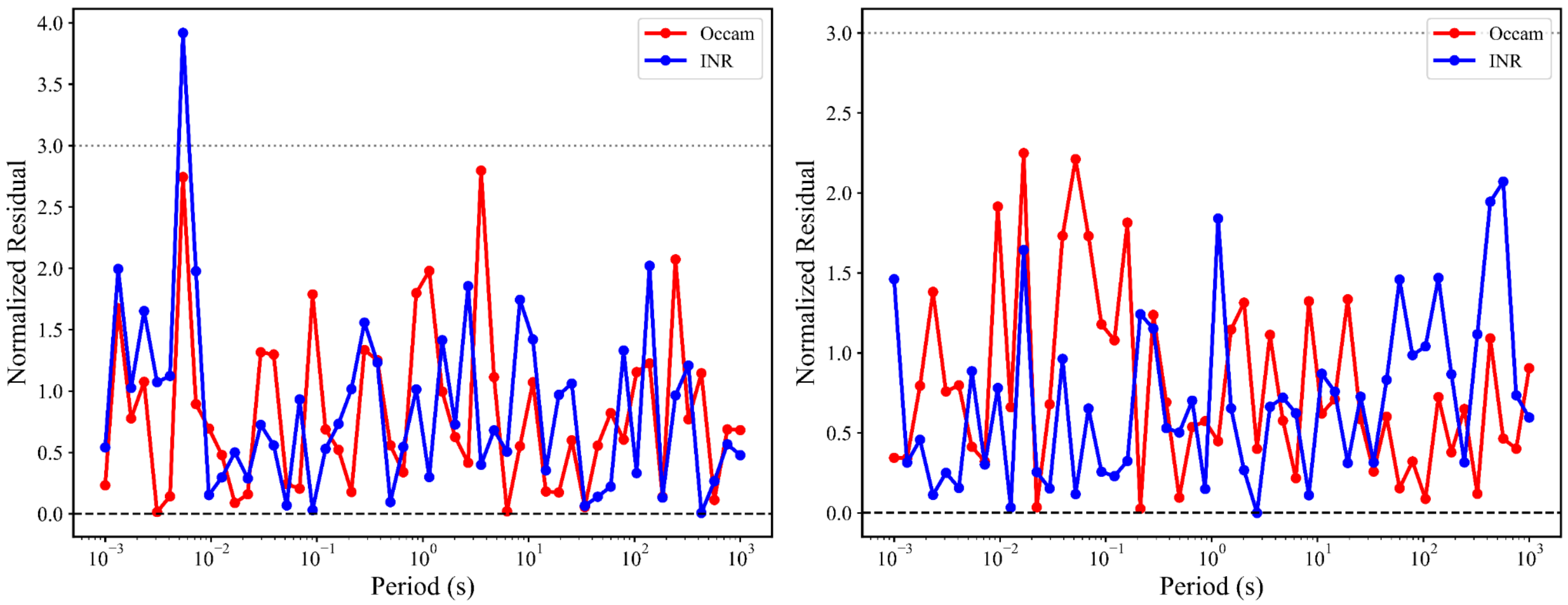

*Fig. 10. Comparison of normalized residuals obtained from the INR and conventional Occam inversion methods for apparent resistivity (left) and phase (right) responses. The residual distributions illustrate the relative data-fitting performance of the two approaches across the investigated period range.*

## 3.2. Real Data Case:

For the application to real data, we used the COPROD magnetotelluric data set from (Constable et al., 1987). We inverted the dataset using the proposed framework of INR inversion. The primary model control variables for this framework are the depth sampling grid and the normalized depth coordinate, $z_{norm}$. These parameters are important for the spatial representation of the subsurface resistivity model and are key for the inversion stability, convergence behavior, and depth resolution.

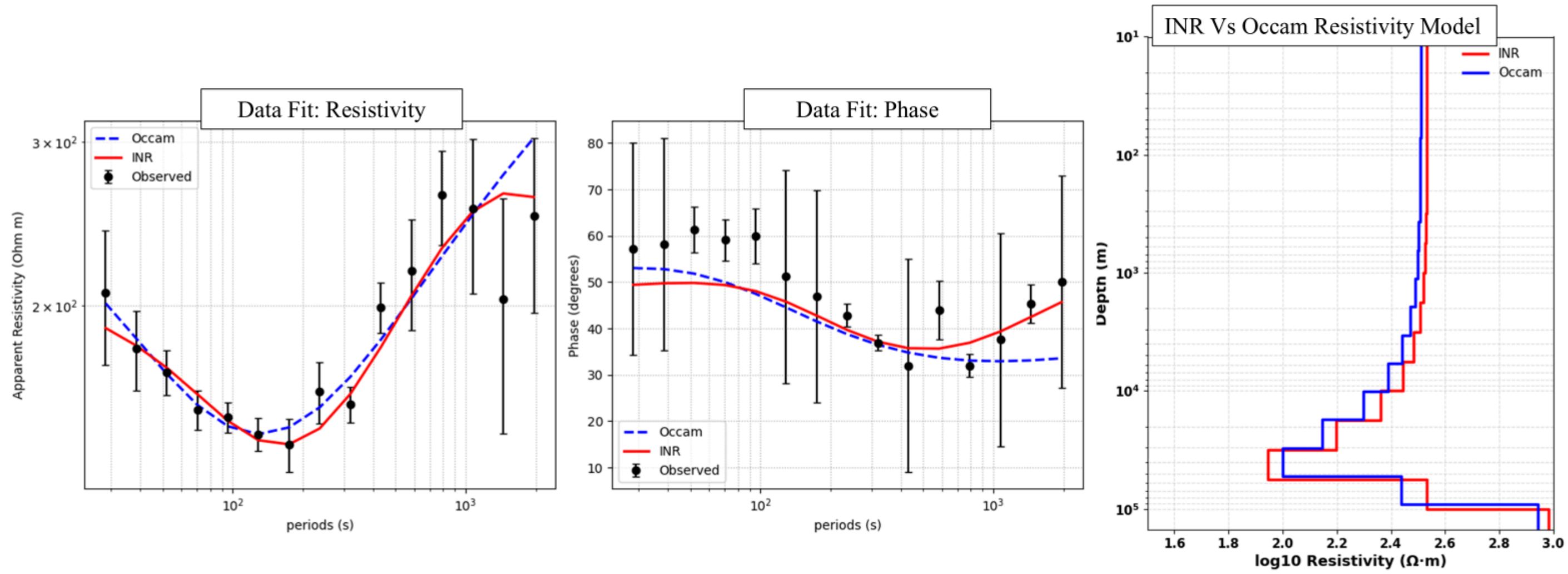

*Fig. 11. Comparison of data fits (left & center) and model responses (right) of INR and OCCAM using COPROD dataset*

We compared the results with the conventional Occam inversion approach to assess the performance of the INR method. In terms of the results obtained from the synthetic experiments, the INR inversion performs better in terms of data fitting for larger depths than the Occam inversion (Fig. 11, left & center). The recovered INR and Occam models follow a consistent resistivity trend with depth; however, differences appear in the transition zones and deeper regions (Fig. 11, right). The smoother nature of the Occam model reflects its regularization strategy, whereas the INR model captures relatively sharper variations.

Normalized residuals of apparent resistivity (Fig. 12a, left) and phase (Fig. 12a, right) were then calculated for the results of the INR and the Occam inversion. The analysis shows a clear peak in the apparent resistivity residuals of the INR inversion at shallow depths, and the residuals at deeper depths are generally smaller than those from the Occam inversion. The same tendency was observed in synthetic experiments, demonstrating consistent behavior of the INR method on synthetic and real data sets. Furthermore, we examined the RMS convergence (Fig. 12b, left) and uncertainty quantification using ensemble inversion with 40 independent runs (Fig. 12b, right).

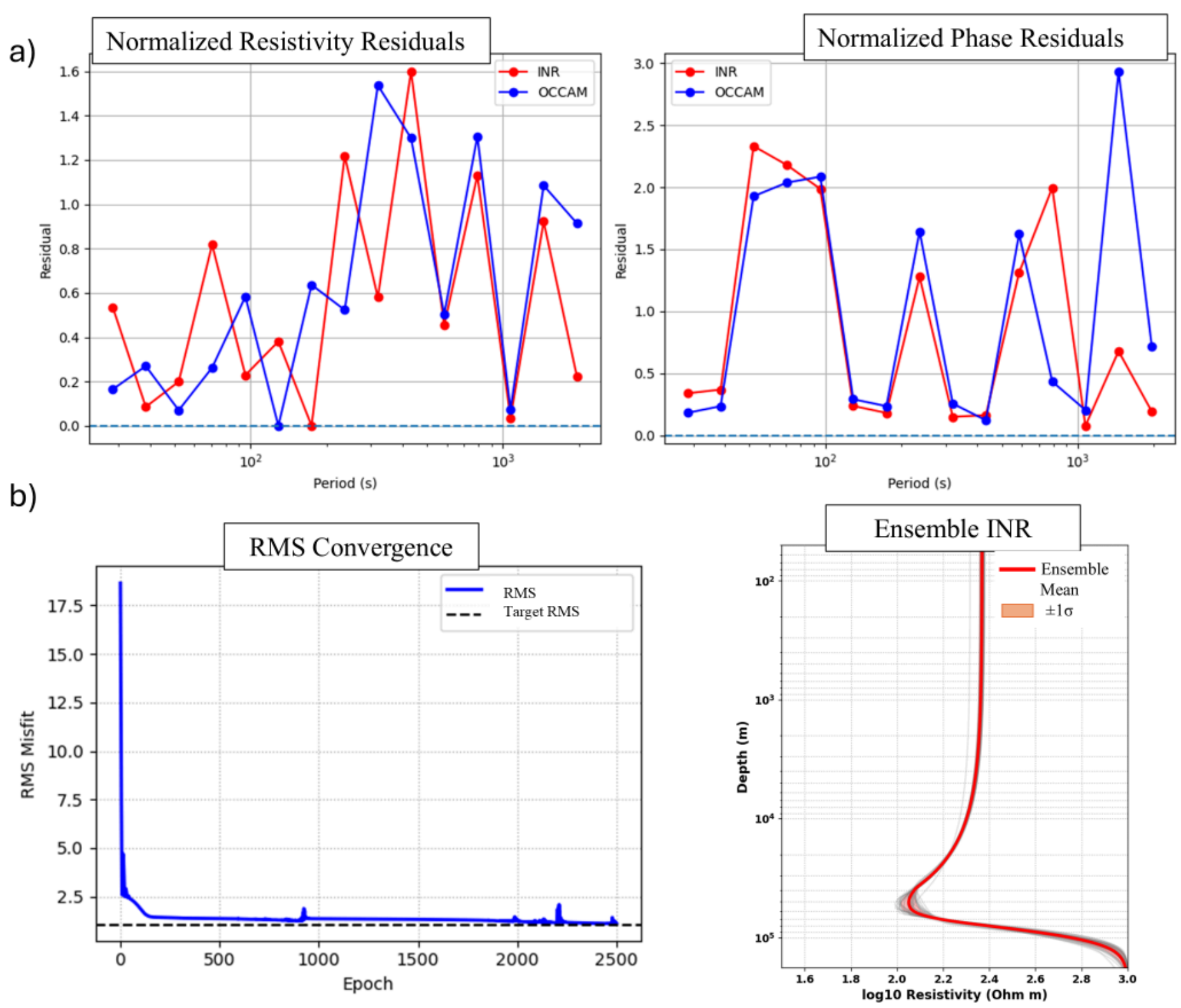

*Fig. 12.* The top panel (a) shows *a comparison of normalized residuals of apparent resistivity (left) and phase (right) obtained from the INR and conventional Occam inversion methods using the COPROD MT dataset. The bottom panel (b) shows RMS convergence (left) and uncertainty quantification results (right) for the INR inversion framework*

Finally, the performance of the INR and Occam inversion methods was evaluated using various other quantitative metrics. These were common error and evaluation measures like mean square error (MSE), mean absolute error (MAE), root mean square error (RMSE), and coefficient of determination ($R^2$) for correlation analysis. Additionally, the Pearson correlation coefficient (r) was used to assess the linear correlation between the observed and predicted responses, and the Structural Similarity Index (SSIM) was used to evaluate the structural similarity between the models. The Correlation Index (CI) was also calculated as another measure of the model consistency and reliability (Saibi et al., 2026).

The INR method gave better results than the Occam inversion in several evaluation metrics. In particular, INR achieved the lowest values of Mean Square Error (MSE=22.9898), Mean Absolute Error (MAE=2.9522), Root Mean Square Error (RMSE=4.7948), and Correlation Index (CI=1.5257). Moreover, it exhibited the highest coefficient of determination ($R^2$=0.9591), Pearson correlation coefficient (r=0.9901), and Structural Similarity Index (SSIM=0.9776).

The high SSIM value indicates that the INR model is able to maintain the vertical layering patterns of the subsurface while resolving the finer structural details. The lesser value of CI indicates better stability of the model with less scattering in the predicted responses. Moreover, the near-unity value shows a good agreement between the observed and predicted data with low variance, and the high Pearson correlation coefficient shows a strong positive linear relationship between the observed and predicted MT responses. Overall, these results suggest that the INR framework describes the subsurface resistivity structure more accurately and reliably than the conventional Occam inversion approach.

Total MT Metrics ($\rho_a + \varphi$)

| Metric | Occam | INR |
| --- | --- | --- |
| MSE | 34.1443 | 22.9898 |
| MAE | 3.5059 | 2.9528 |
| RMSE | 5.8433 | 4.7948 |
| $R^2$ | 0.9393 | 0.9591 |
| Correlation | 0.9882 | 0.9901 |
| 95% CI | 1.7422 | 1.5257 |
| SSIM | 0.9663 | 0.9776 |

*Table 2: Performance metrics for INR  and Occam inversion on COPROD MT dataset*

## 4. Conclusion

In this work, a physics-driven implicit neural representation (INR) framework for 1D magnetotelluric inversion has been developed. The subsurface resistivity is modeled as a continuous neural function of depth, thereby avoiding the need for fixed layer discretization and enabling flexible neural-field-based model parameterization. An automatic differentiation scheme is used to embed a differentiable MT forward solver based on Wait's recursive formulation to perform a fully physics-consistent inversion without explicit calculation of sensitivities. In contrast to traditional approaches, which require extensive manual tuning of regularization parameters, the implicit regularization is provided by the architecture of the neural network and training dynamics. This approach avoids the dependence on user-supplied smoothing constraints while retaining, as far as possible, the important structural features of the subsurface.

Synthetic experiments show the INR framework can recover resistivity structures of different depths and thicknesses. The resolution for thin conductive bodies, however, decreases at greater depths because of the inherent limitations of MT sensitivity. The stability of the inversion is further confirmed by an ensemble-based uncertainty quantification with multiple random initializations that shows the recovered models are subject to very little variability.

As compared with Occam inversion, the INR method has consistently better fitting of the data, especially in the deeper part. It also outperforms in several quantitative measures such as MSE, RMSE, $R^2$, correlation coefficient, CI, and SSIM. This method also reduces residuals in the deeper parts, which means that the model is more stable and reliable. Experiments on real MT data provide additional evidence for the effectiveness of the framework, showing consistent behavior with synthetic experiments as well as the applicability of the framework in practice.

**Code Availability Statement**

The codes for reproducing the results will be open upon acceptance of the manuscript.

**Acknowledgement**

This research was financially supported by DST-INSPIRE [Research Fellowship to FBS (IF No. 220411)]  from the Government of India. The authors would like to acknowledge the Indian Institute of Petroleum and Energy, Visakhapatnam, for providing necessary research facilities and support.

**Declaration of AI usage**

GPT-5.5 (OpenAI) was used in a limited role during manuscript polishing and code cleanup and structuring, primarily for language editing, formatting, and minor text refinement. All methodological decisions, analyses, implementations, and conclusions were carried out and verified by the authors

**Author Contributions**

Conceptualisation: FBS, PKM; Code development: FBS, PKM, RKS; Numerical experiments: FBS, PKM; Validation: FBS, RKS; Investigation and Supervision: RKS; Manuscript writing and editing: FBS, RKS; All authors reviewed the manuscript.